\documentclass[a4paper,fleqn]{cas-sc}
\usepackage{caption}
\usepackage{float}

\usepackage[numbers]{natbib}

\usepackage[table]{xcolor}
\definecolor{rowgray}{gray}{0.93}

\def\tsc#1{\csdef{#1}{\textsc{\lowercase{#1}}\xspace}}
\tsc{WGM}
\tsc{QE}
\begin{document}
\let\WriteBookmarks\relax
\def\floatpagepagefraction{1}
\def\textpagefraction{.001}

\shorttitle{Security, Privacy, and Ethical Risks in OpenClaw}    

%\shortauthors{Yutong Jin et al.}  

\title [mode = title]{Security, Privacy, and Ethical Risks in OpenClaw}  



\author[1]{Yutong Jin}
\fnmark[1]
\ead{yutong.jin@queensu.ca} 
\ead[url]{}
\credit{}

\author[1]{Zelin Zhang}
\fnmark[2]
\ead{zelin.zhang@queensu.ca}
\ead[url]{}

\author[2]{Zhijin Lyu}
\fnmark[3]
\ead{25bm38@queensu.ca}
\ead[url]{}

% Corresponding author
\author[1]{Jianbing Ni}
\cormark[1]
\fnmark[4]
\ead{Jianbing.ni@queensu.ca}
\ead[url]{}

% Shared affiliation
\affiliation[1]{organization={Department of Electrical and Computer Engineering, Queen's University},
 %          city={Kingston},
            postcode={K7L 3N6}, 
            country={Canada}}

            \affiliation[2]{organization={Department of Mechanical and Materials Engineering, Queen's University},
            city={Kingston},
            postcode={K7L 3N6}, 
            country={Canada}}

\cortext[1]{corresponding author}
%\fntext[2]{}
%\fntext[3]{}
%\fntext[4]{}


\begin{abstract}
This paper systematically investigates the security, privacy, and ethical risks, as well as the traceability challenges of OpenClaw, a locally executable AI agent system for natural language interaction and real-world task completion. While OpenClaw shows strong potential for personal assistance, office automation, cross-platform task management, and information integration, it also raises serious security, privacy, and ethical concerns. By analyzing its system architecture, core functionalities, deployment model, and representative application scenarios, this paper aims to reveal the risks that may arise when such a highly privileged agent is integrated into personal and organizational digital environments. We focus in particular on the challenges associated with persistent local storage, tool invocation, cross-context information aggregation, multi-user interaction, and the integration of plugins and external services. We argue that these issues constitute major barriers to the trustworthy deployment and widespread adoption of this technology. Finally, we summarize the open challenges in security defenses, privacy protection, ethical governance, and traceability in agent use, and call for joint efforts from researchers, developers, deployers, and regulators to build AI agent systems that are safer, more reliable, and more trustworthy.
\end{abstract}


%\item 
%\item 
%\item 

% Keywords


\maketitle

\section{Introduction}
Recent advances in large language models (LLMs) have enabled the emergence of autonomous AI agents capable of executing complex tasks with minimal human supervision \cite{shu2024towards}. Unlike traditional conversational systems, autonomous agents can go beyond response generation by invoking tools, accessing external resources, and maintaining state across interactions \cite{schick2023toolformer, chase2022langchain}. OpenClaw \cite{steinberger2026openclaw} exemplifies this shift as an open-source platform that connects LLMs with persistent sessions, browser interaction, messaging channels, and extensible skills. The OpenClaw agent is not limited to answering questions. It can search and summarize information from external sources, send and receive messages across communication channels, open webpages and carry out browser-based tasks, maintain context across long-running sessions, and call specialized tools or skills to complete multi-step workflows. Compared with earlier LLM-based systems that mainly focused on constrained text generation or task-specific natural language processing, platforms such as OpenClaw move language models from passive text generation toward action-oriented digital assistance. This shift substantially broadens the potential applications of AI systems in personal assistance, workflow automation, communication, and productivity support, while also making OpenClaw a useful case study for understanding the broader development of agent-based AI platforms.

However, the capabilities of AI agents introduce new security, privacy, and ethical concerns. An agent that can read files, browse webpages, invoke tools, and send messages can also be manipulated into performing actions that ordinary conversational models could never directly carry out. For example, content controlled by an attacker may shape the agent’s interpretation of a task, and browser-based workflows may expose local files or authenticated user sessions; persistent transcripts and session state may retain sensitive information beyond a single interaction; and poorly designed or malicious skills may expand the system's behavior in unsafe ways. In addition, once such platforms are deployed in real communication and workflow settings, their risks are no longer limited to incorrect text generation. They can affect privacy through unintended disclosure, affect security through unsafe actions on local or remote resources, and raise broader concerns about trust, accountability, transparency, and the responsible use of AI-driven automation \cite{su2025survey,deng2025ai,narajala2025securing,lupinacci2025dark}. Therefore, a careful identification and structured analysis of the risks of platforms such as OpenClaw is necessary, both to improve risk awareness in real-world use and to support the design of more effective safeguards, ultimately enabling more trustworthy deployment of such systems.

In this paper, we examine OpenClaw, a recent autonomous AI agent designed for natural language interaction and real-world task execution, through the analytical framework shown in Fig.~\ref{fig:openclaw-overview}. Specifically, we organize the discussion around four closely related dimensions: security, privacy, ethics, and traceability. This structure allows us to connect OpenClaw's architectural design with the risks, impacts, and safeguards associated with execution-capable agent systems. Our aim is to draw greater attention to these challenges and to clarify directions for building safer and more trustworthy AI agents. The main contributions of this paper are as follows.

\begin{center}
    \includegraphics[width=0.75\linewidth]{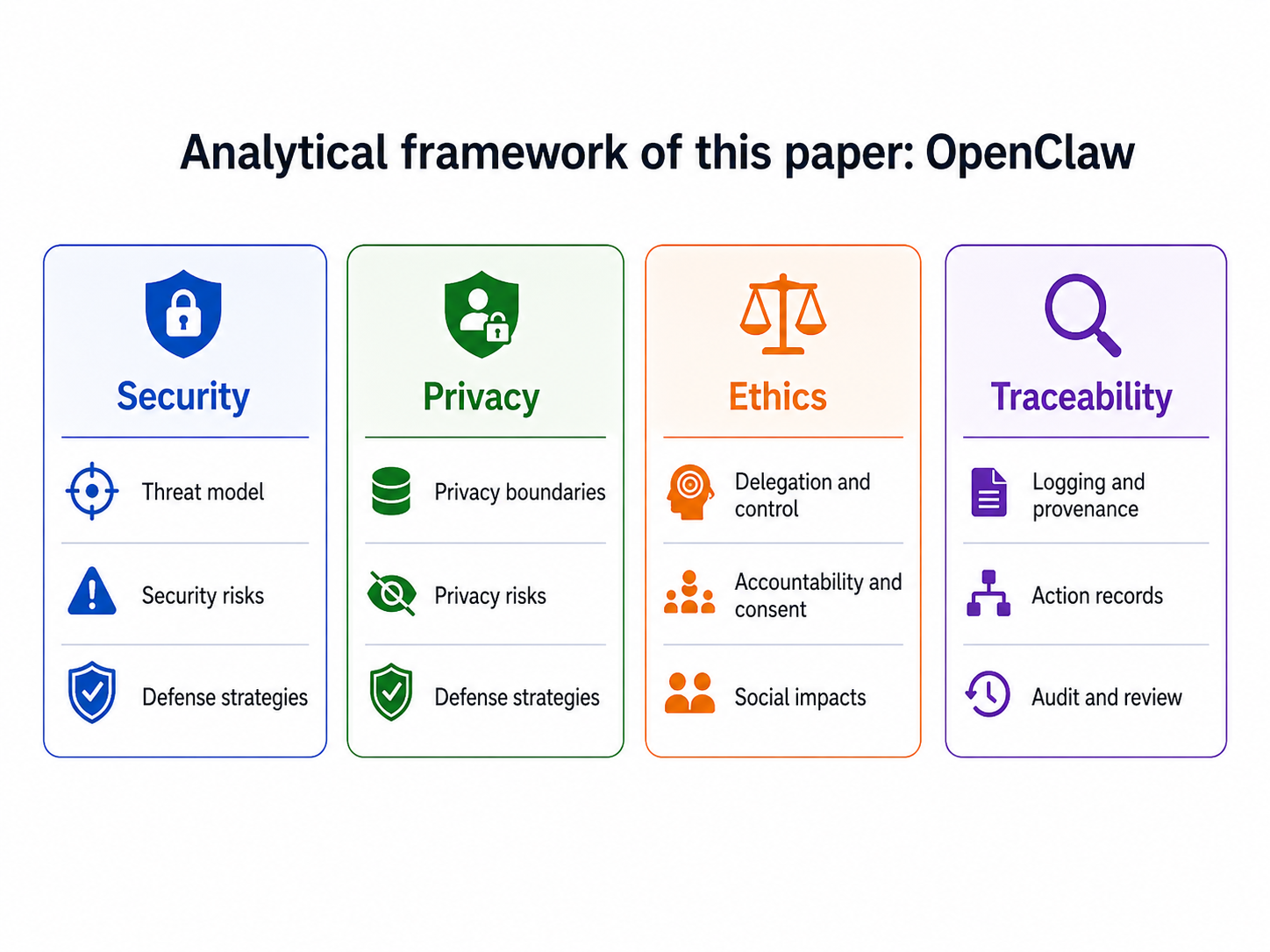}
    \captionof{figure}{Analytical framework of the paper for OpenClaw.}
    \label{fig:openclaw-overview}
\end{center}

\begin{itemize}

\item We present a structured overview of OpenClaw as an autonomous AI agent platform, covering both the general agent capabilities of planning, memory, and tool use, and the specific design of its self-hosted gateway, embedded runtime, tool layer, skill mechanism, and persistent sessions. We also discuss representative applications and show how these capabilities introduce new trust boundaries together with practical utility.

\item We analyze the principal security risks of OpenClaw across its gateway, runtime, tool layer, skill ecosystem, and persistent session state, including indirect prompt injection through external content ingestion, unsafe and over-privileged tool execution, browser risks from authenticated sessions, supply-chain vulnerabilities in third-party skills, and cross-session contagion enabled by persistent state and weak trust boundaries.
    
\item We discuss the major privacy issues raised by OpenClaw, including broadened privacy boundaries, persistent memory and workspace storage, over-broad access to tools and accounts, leakage across users and contexts, and additional privacy risks introduced by plugins, untrusted external content, and insufficient auditing.

    \item We analyze the broader ethical concerns associated with OpenClaw and similar agent systems, including accountability, transparency, trust, misuse, and the social consequences of increasingly capable AI assistants.
    
    \item We outline the open challenges and future research directions for improving the security, privacy, and trustworthiness of OpenClaw like autonomous agent platforms.
\end{itemize}

\section{OpenClaw}
\subsection{Autonomous AI Agent Architecture}

Autonomous AI agents are commonly built around a LLM
that serves as the core reasoning engine. In contrast to traditional
conversational systems, an autonomous agent is not limited to generating
textual responses; it can plan, retain and retrieve contextual information,
and interact with external tools or environments in order to achieve a user
goal \cite{yao2023react,schick2023toolformer,park2023generative}.

A common conceptual view of LLM-powered agents is to consider three major
capabilities that enable autonomous behavior: planning, memory, and tool use \cite{weng2023agents}. This decomposition has become a
common conceptual framework for describing modern autonomous agents.

\paragraph{Planning.}
Planning refers to the agent's ability to decompose a complex task into
intermediate steps, select actions, and revise its strategy over multiple
iterations. In many agent systems, this capability is implemented through
a reasoning--action loop, where the model alternates between generating
internal reasoning traces and taking actions in the external environment.
Representative examples include ReAct-style trajectories, where the model
iteratively produces \texttt{Thought}, \texttt{Action}, and
\texttt{Observation} sequences \cite{yao2023react}. More advanced agent
architectures may also incorporate self-reflection or refinement mechanisms
to improve task execution over time.

\paragraph{Memory.}
Memory allows the agent to retain and retrieve information beyond a single prompt window. Agent memory can be organized at different temporal scales. Short-term memory corresponds to information available within the current context window, while longer-term memory is often implemented through external storage mechanisms such as vector databases or persistent files \cite{weng2023agents,park2023generative}. Memory is particularly important for long-running tasks, contextual continuity, and experience accumulation across sessions.

\paragraph{Tool Use.}
Tool use refers to the ability of an agent to invoke external functions, APIs, or execution environments in order to obtain information or perform actions that are not directly achievable through language generation alone. Examples include web retrieval, code execution, access to proprietary data sources, and interactions with external software systems. Tool-augmented language models have been shown to significantly improve task-solving capabilities in realistic environments \cite{schick2023toolformer,yao2023react}.

Overall, planning, memory, and tool use transform an LLM from a passive text generator into an active task-oriented system. Meanwhile, the integration of these capabilities introduces multiple trust boundaries between the models, external resources, and persistent state, which motivates the security analysis presented in the following sections.

\subsection{OpenClaw System Architecture}

OpenClaw is an open-source autonomous AI agent platform built around a self-hosted gateway architecture \cite{steinberger2026openclaw}. Unlike a conventional chat system that mainly provides a stateless conversational interface, OpenClaw is designed as a long-running system that manages communication, session continuity, access control, and agent execution within a unified framework. According to its official documentation, the platform can connect to multiple communication channels, including WhatsApp, Telegram, Slack, Discord, Signal, iMessage, and WebChat, while also supporting local and remote control clients.

From a system perspective, OpenClaw can be understood through five closely related components: the gateway, the agent runtime, the tool layer, the skill mechanism, and persistent sessions with stored state. Beyond describing what each component does, it is also important to note that these components define where trust boundaries are created and how risks may later arise in practice.

\begin{center}
    \includegraphics[width=0.60\linewidth]{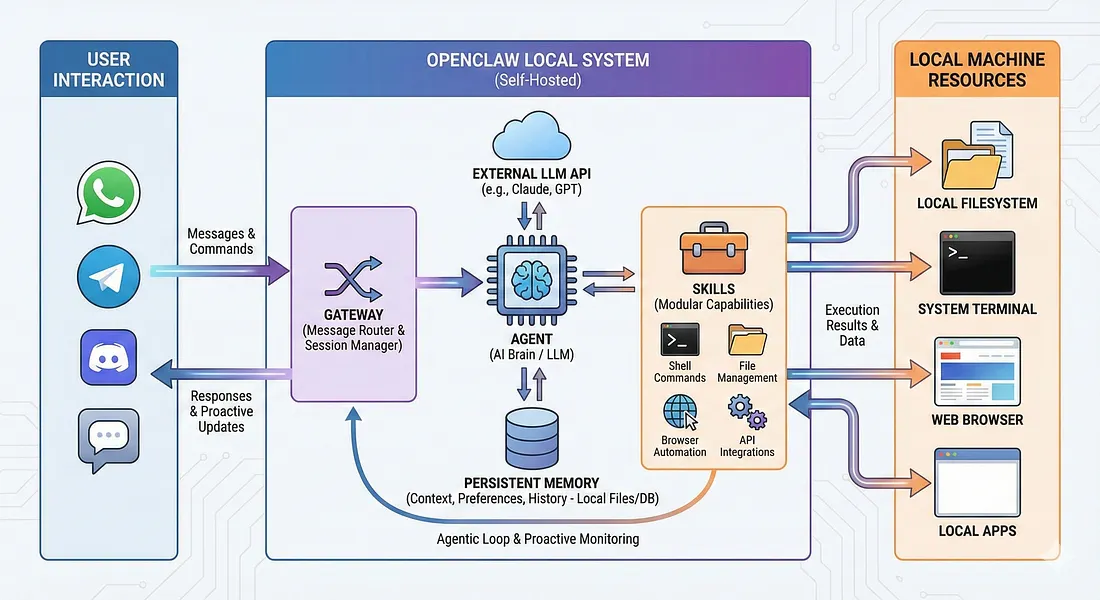}
    \captionof{figure}{System architecture and main functions of OpenClaw. Redrawn based on OpenClaw's official architecture documentation and a community-circulated explanatory diagram.}
    \label{fig:openclaw_structure}
\end{center}

\paragraph{Gateway Layer.}
The gateway serves as the central coordination point of the platform. It maintains channel connections, receives incoming events, routes requests into active sessions, and returns outputs to external interfaces. OpenClaw can also attach optional device nodes to the same gateway through WebSocket, with each node declaring its role and available capabilities. As a result, the gateway functions not only as a routing hub, but also as a control layer for pairing, authentication, access management, and other system-level decisions. This centralization is functionally useful, but it also makes the gateway a key trust boundary, since channel access, routing logic, and session association all affect how information and influence can enter and propagate through the system.

\paragraph{Agent Runtime.}
At the core of OpenClaw is an embedded runtime that combines language model generation with a structured reasoning and action process. Rather than treating the model as an isolated conversational endpoint, OpenClaw manages the full session lifecycle within its own runtime. This includes prompt construction, event handling, tool injection, streaming, and session updates. User inputs therefore do not simply produce text replies. They may also trigger tool calls, state changes, and external actions. Because the runtime is responsible for assembling context and mediating between model outputs and tool-enabled behavior, it is also a key point at which untrusted content, prior state, and retrieved information may shape later reasoning and execution.

\paragraph{Tool Layer.}
OpenClaw includes a structured tool system that allows the agent to interact with external resources and digital environments. These tools may involve file access, browser operations, messaging functions, session utilities, automation support, and other platform-specific capabilities. Through this layer, model outputs can be translated into actions that affect files, services, devices, or web content. This is one of the main features that distinguishes OpenClaw from a standard chatbot. At the same time, it also makes the tool layer a major risk-amplification point, because errors in reasoning, unsafe context interpretation, or injected instructions may be converted into real actions on local or remote systems.

\paragraph{Skill Mechanism.}
Another important part of the platform is its skill mechanism, which allows reusable capabilities to be added without modifying the core runtime. A skill typically includes a \texttt{SKILL.md} file that defines its intended use and may also contain supporting metadata or integration logic. This design improves extensibility and task adaptability, but it also broadens the operational surface of the platform and introduces additional points of risk. In particular, third-party skills expand the system's trust boundary by introducing external logic and capabilities that may influence reasoning, permissions, and execution behavior.

\paragraph{Persistent Sessions and State.}
OpenClaw also differs from ordinary prompt-response systems in its use of persistent sessions, transcripts, and task-related artifacts. Session keys support continuity across interactions, while stored state allows information from earlier exchanges to shape later behavior. This persistence is valuable for long-running and multi-step tasks, but it also increases the importance of state management, data handling, and access control. More importantly, it means that errors, injected content, or sensitive information may survive beyond a single interaction and continue to affect later behavior across time, channels, or users.

Taken together, these components make OpenClaw a stateful agent system that can participate directly in digital workflows rather than merely generate text. More importantly, they also define the main trust boundaries through which risk arises. The gateway concentrates cross-channel access and routing authority, the runtime constructs the context that shapes reasoning and action, the tool layer converts model outputs into real-world effects, the skill mechanism extends the platform through third-party logic, and persistent sessions allow information and influence to carry across interactions. For this reason, OpenClaw's architecture is not only a functional design, but also the basis of its later security, privacy, and governance risk profile.

\subsection{Representative Applications}
Communication and information coordination assistant: The agent can access and operate across multiple messaging channels while maintaining persistent context over time. It can retrieve, aggregate, and process incoming messages from platforms such as Slack or WhatsApp, and assist users in managing ongoing communication based on prior interactions stored in the session. For example, a user could request “summarize my recent conversations and generate a to-do list.” The agent can then extract key points from messages, identify actionable items, and produce a structured to-do list. In more advanced scenarios, it may further execute actionable items from the list, such as scheduling meetings, sending follow-up messages, or creating tasks in external tools. As a result, OpenClaw can function as a unified coordination layer, reducing the need for users to manually track action items across conversations and applications.

Information aggregation and retrieval: The agent integrates data from multiple sources and produces structured outputs. Unlike conventional systems that rely on isolated queries, the agent can combine information from web resources, local files, prior conversations, and user preferences to support more context-aware results. For instance, a user may ask the agent to “recommend a refrigerator for a small apartment within a given budget.” The agent can retrieve product information from online sources, incorporate constraints such as space, budget, and brand preferences, and generate a structured comparison of suitable options.In more advanced cases, it may further cluster products by use scenarios, align recommendations with user profiles or demand patterns, and prepare procurement plans across multiple units or locations. This capability extends beyond individual query-based retrieval by enabling cross-source integration, constraint-aware filtering, and structured decision support within a continuous interaction context.

Technical task execution and iteration: persistent context and tool access enable the agent to carry out and refine technical tasks over time. For example, a user may ask the agent to “set up a comparative benchmark of object detection methods on my dataset.” The agent can retrieve relevant implementations, standardize input formats, and prepare a consistent preprocessing pipeline tailored to the dataset. It may then generate initial configurations and scripts to run multiple methods under comparable settings, and produce structured summaries of results. In more advanced cases, it may further refine implementations by reorganizing modules, adjusting configurations, or integrating improved components based on observed performance. By maintaining continuity across iterations and coordinating actions over multiple tools, OpenClaw supports the efficient execution and refinement of dataset-specific technical workflows without requiring repeated manual setup.

These examples illustrate the practical value of OpenClaw, but they also clarify why the platform deserves closer scrutiny. Because it is designed not only to respond, but also to act, store context, and interact with external systems, its deployment raises broader questions about security, privacy, and trust. These issues are examined in the following sections.
\begin{center}
    \includegraphics[width=0.80\linewidth]{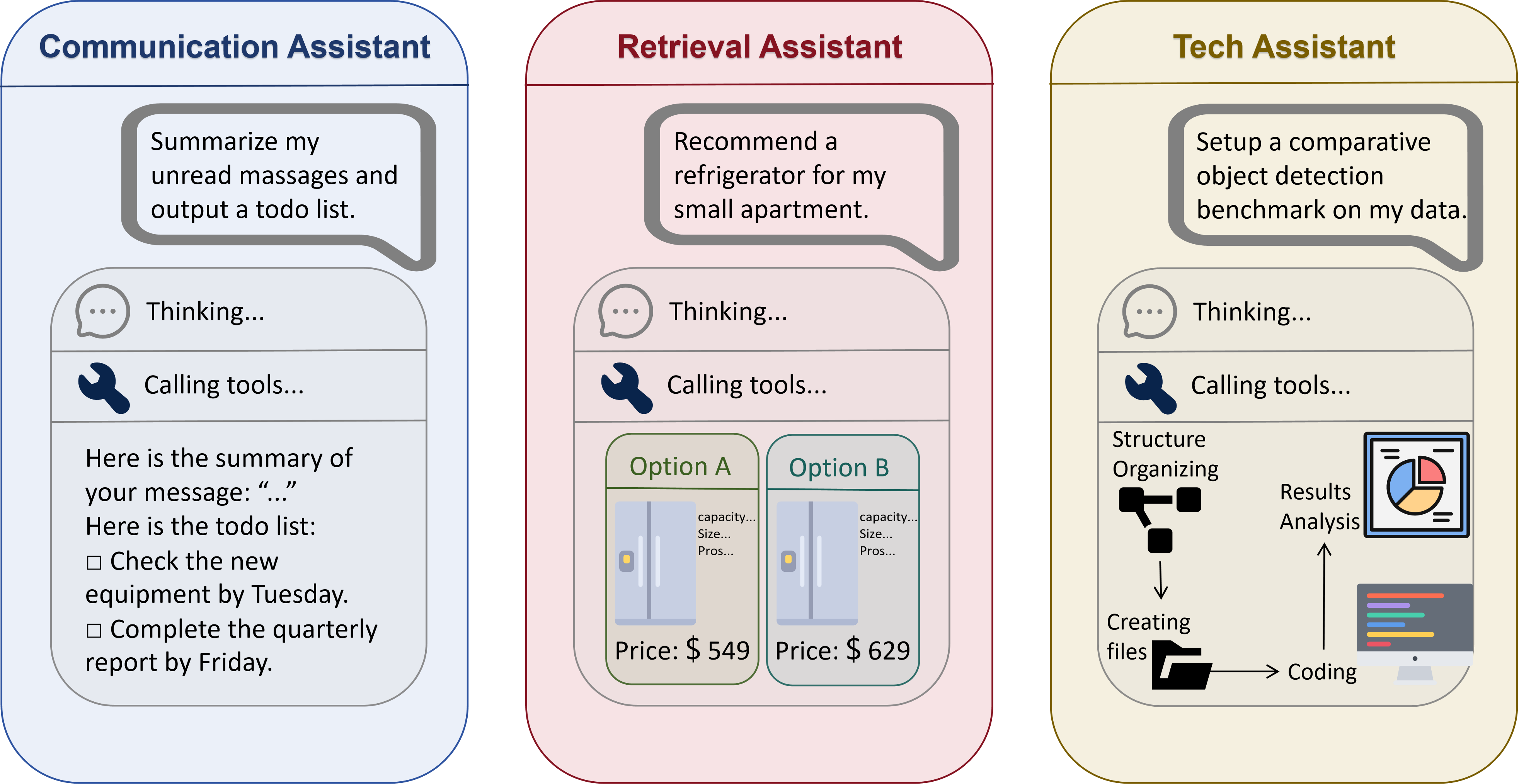}
    \captionof{figure}{Representative application scenarios of OpenClaw: communication, retrieval, and technical assistance}
    \label{fig:openclaw_application}
\end{center}

\section{OpenClaw vs. Security}

\subsection{Threat Model}

OpenClaw's security analysis requires a threat model that reflects both its architecture and its deployment conditions. Unlike conversational chatbots that operate only on text inputs and outputs, OpenClaw exposes a multi-layered attack surface through its gateway, runtime, tool layer, skill mechanism, and persistent session state. The severity of these risks depends on how the platform is deployed, for example whether it is used by a single user or shared by multiple users, whether it is reachable through external channels, whether third-party skills are enabled, and whether it is connected to sensitive accounts or authenticated browser sessions. This section therefore defines the threat model along three dimensions: adversary identity, adversary capability, and adversary objective.

\subsubsection{Adversary Identities and Attack Entry Points.}
Adversaries may initiate attacks from multiple layers of the OpenClaw architecture. At the channel layer, the adversary may be an external user who sends malicious inputs through messaging channels such as WhatsApp, Telegram, or Slack, or a participant in a public channel who posts content containing hidden instructions. At the content layer, the adversary may control third-party content such as webpages, documents, or emails that OpenClaw retrieves during task execution. At the skill layer, the adversary may be a developer who distributes malicious skills through ClawHub. At the device layer, the adversary may attempt to impersonate a legitimate device node and connect to the gateway through the WebSocket interface. In shared deployments, another user of the same system may also become a source of cross-session or cross-context influence.

\subsubsection{Adversary Capabilities.}
The adversary's core capabilities include injecting adversarial instructions into content that enters the OpenClaw context window, achieving persistent influence by poisoning session state or retrieved memory, exploiting trust relationships between sessions, and expanding the platform's attack surface by uploading skills that contain malicious logic. These capabilities depend on concrete preconditions. For example, indirect injection requires control of content that the agent is likely to ingest, skill-based attacks require a path for extension installation or activation, and session-contamination attacks require shared state, weak isolation, or later access to a context shaped by earlier malicious input.

\subsubsection{Adversary Objectives.}
The adversary's objectives can be grouped into three broad categories: \textit{goal hijacking}, that is, redirecting the agent's execution flow so that it performs operations specified by the adversary rather than those intended by the user; \textit{data exfiltration}, that is, extracting conversation history, local file contents, user credentials, or other sensitive contextual information; and \textit{destructive execution}, that is, triggering harmful tool-mediated actions such as file deletion, unauthorized message sending, or malicious code execution.

Several threat modeling frameworks have provided systematic analyses of agent security. Narajala and Narayan~\cite{narajala2025securing} proposed a taxonomy of nine threat categories across five domains, with each category mapped to the STRIDE framework. Su et al.~\cite{su2025survey} further argued that the risks facing agent systems include fundamentally new threats arising from agent autonomy, persistent state, and the capacity for irreversible actions, beyond known vulnerabilities inherited from the underlying LLM. Crucially, the research community has already distinguished between \textit{threats to the agent} and \textit{threats through the agent}~\cite{deng2025ai,su2025survey}. As autonomy increases, the agent shifts from a vulnerable attack target to a powerful attack vector, a transition clearly reflected in the architecture of OpenClaw. In practice, the relative importance of these risks varies across deployment settings, with stronger cross-user risks in shared instances, stronger remote-input risks in internet-facing deployments, and stronger exfiltration and destructive-execution risks when sensitive accounts or high-privilege tools are enabled. To make this threat model more concrete, Table~\ref{tab:attack_scenarios} summarizes representative attack scenarios in OpenClaw and the architectural components they primarily target.

\begin{table*}[htbp]
\centering
\caption{Representative attack scenarios against OpenClaw by risk category.}
\label{tab:attack_scenarios}
\fontsize{7}{7}\selectfont
\setlength{\tabcolsep}{6pt}
\renewcommand{\arraystretch}{1.3}
\begin{tabular}{l l l}
\toprule
\textbf{Risk Category} & \textbf{Attack Type} & \textbf{Target Component} \\
\midrule
\multirow{4}{*}{Indirect Injection}
  & \cellcolor{rowgray}Channel message injection
  & \cellcolor{rowgray}Gateway \\
  & Web content poisoning
  & Browser tool \\
  & \cellcolor{rowgray}Document-embedded payload
  & \cellcolor{rowgray}Tool layer \\
  & Encoded instruction delivery
  & Runtime \\
\midrule
\multirow{4}{*}{Tool and Permission}
  & \cellcolor{rowgray}Path traversal escape
  & \cellcolor{rowgray}File tool \\
  & Cross-tool composition
  & Browser + File tool \\
  & \cellcolor{rowgray}Unconfirmed irreversible action
  & \cellcolor{rowgray}Tool layer \\
  & Authenticated session hijack
  & Browser tool \\
\midrule
\multirow{3}{*}{Skill Supply Chain}
  & \cellcolor{rowgray}Malicious skill via ClawHub
  & \cellcolor{rowgray}Skill mechanism \\
  & MCP protocol exploitation
  & Skill mechanism \\
  & \cellcolor{rowgray}Silent breakage from refactoring
  & \cellcolor{rowgray}Skill interface \\
\midrule
\multirow{3}{*}{Multi-Session Contagion}
  & Persistent state poisoning
  & Session store \\
  & \cellcolor{rowgray}Cross-channel propagation
  & \cellcolor{rowgray}Gateway \\
  & Cross-user chain propagation
  & Gateway + Messaging \\
\bottomrule
\end{tabular}
\end{table*}

\subsection{Security Risks}

This section examines the principal security risks of OpenClaw, organized by the architectural component from which each risk primarily arises.

\subsubsection{External Content Ingestion and Indirect Prompt Injection}

OpenClaw is designed in a way that requires it to process large amounts of content from untrusted sources during normal workflows. In a typical use case, such as ``summarize my recent Slack conversations and generate a to-do list,'' the agent needs to retrieve content from multiple messaging channels, gather information from the web, read local documents, and reason over all of this content within a single context window. At the architectural level, this process passes through the gateway layer, the agent runtime, and the tool layer. This creates ideal conditions for indirect prompt injection.

The key difference between indirect prompt injection~\cite{greshake2023not} and direct prompt injection is that the attacker does not need to interact with the agent directly. Instead, the attacker places malicious instructions in third-party content that the agent is likely to retrieve during task execution, such as web pages, shared documents, messages in communication channels, email attachments, or database entries. When the agent retrieves and processes this content during normal operation, the malicious instructions become part of the context and may alter the agent's understanding of the task and its subsequent actions.

For OpenClaw, the risk of indirect injection is significantly amplified by three architectural features. First, OpenClaw processes untrusted input from multiple messaging channels and web sources at the same time. Second, OpenClaw has direct access to sensitive systems, including the file system, the shell, APIs, and authenticated web sessions. Third, the tool layer allows it to produce irreversible state changes, such as sending messages, executing code, and modifying or deleting files. The simultaneous presence of these three conditions violates the so-called Rule of Two in agent security, which states that an agent should satisfy at most two of the following three conditions: processing untrusted input, accessing sensitive systems, and producing external state changes~\cite{debenedetti2025camel}. OpenClaw satisfies all three by design.

Consider a concrete attack path. A user asks OpenClaw to summarize recent Slack conversations. An attacker posts an apparently normal message in a public Slack channel that the user belongs to, embedding a hidden instruction through Unicode control characters or visually indistinguishable formatting. When OpenClaw retrieves messages from that channel and constructs the context during runtime, the embedded instruction enters the reasoning process. This may cause the agent to forward the conversation summary to a channel chosen by the attacker, or to embed content from local files into an otherwise normal-looking reply, while the to-do list itself appears normal to the user.

Benchmark results confirm that this risk is real. InjecAgent~\cite{zhan2024injecagent} found that GPT-4 using the ReAct paradigm was affected by indirect injection in 24\% of cases under the baseline setting and 47\% under the stronger attack setting. AgentDojo~\cite{debenedetti2024agentdojo} showed that even without adversarial pressure, state-of-the-art LLMs completed fewer than 66\% of tasks correctly, while the most effective attack achieved a targeted success rate of 53.1\%. Researchers from OpenAI, Anthropic, and Google DeepMind jointly evaluated 12 recently proposed defense mechanisms and found that the attack success rate exceeded 90\% against most of them~\cite{nasr2025attacker}. Evaluations of OpenClaw itself reflect the same reality. Chen et al.~\cite{chen2026trajectory} reported that OpenClaw achieved only 57\% prompt injection robustness in a trajectory-based security audit.

\subsubsection{Tool Execution, Permission Control, and Browser Risks}

The tool layer is the key component that turns OpenClaw from a conversational system into an action execution system. Through the tool layer, the output of the language model can be translated into actual operations on the file system, browser, messaging channels, shell, and external APIs. This means that any error in model reasoning, whether caused by prompt injection, misunderstanding of user intent, or the model's own limitations, may directly turn into harmful operations on real digital assets.

The tool layer in OpenClaw is designed to give the agent broad execution authority. The file access tools can read and write files on the local file system, the messaging tools can send messages across multiple channels on behalf of the user, and the browser tools can access web pages and interact with authenticated web sessions. However, the current architecture lacks a fine-grained mechanism for permission separation across tools. A tool invoked to ``read text files in a specified directory'' may, at the architectural level, also have the ability to read system configuration files or access sensitive files in other directories. This permission model violates the principle of least privilege.

Many operations supported by the tool layer are irreversible. A sent message cannot be withdrawn, a deleted file may not be recoverable, and executed code may produce persistent side effects. The security audit by Chen et al.~\cite{chen2026trajectory} found that the pass rate on the intent misunderstanding dimension was 0\%, meaning that every test case involving ambiguity in user intent resulted in an unsafe outcome. The audit also identified an important cascade effect, where a small early misunderstanding of user intent is gradually amplified through later tool calls that change system state. The event documented by Deng et al.~\cite{deng2026taming}, described as Instruction Amnesia via Context Compression, provides a concrete example. The context compression mechanism silently pushed out safety constraints from the system prompt, causing the agent to delete the user's entire email inbox without any confirmation step.

The browser tool introduces additional risks. When the agent accesses a web page, it may inherit the user's existing authenticated session, such as a logged-in email account, bank account, or enterprise intranet. Under indirect injection or misunderstanding of user intent, the agent may be steered into performing actions in these authenticated sessions that the user never intended. Similarly, the combination of browser tools and file tools may allow local assets to be uploaded to external services or embedded into outgoing messages. This cross-tool permission composition means that even if the permissions of each individual tool appear reasonable, coordinated calls across tools may still create an exposure surface beyond what was expected.

Deng et al.~\cite{deng2026taming} systematically evaluated tools contributed by the OpenClaw community and found that about 26\% of them contained security vulnerabilities. Ying et al.~\cite{ying2026uncovering} showed that OpenClaw had a defense success rate of only 17\% in tool-calling scenarios when no human approval step was included, while adding human approval increased the rate to between 19\% and 92\%.

\subsubsection{Skill Ecosystem and Supply Chain Risk}

OpenClaw's skill mechanism allows new capabilities to be added to the agent without modifying the core runtime. Skills are distributed as modules containing a \texttt{SKILL.md} definition file and can be shared through the ClawHub platform. ClawHub already hosts more than 5,700 third-party skills. This extensibility extends the platform's security boundary from its own codebase to the entire skill ecosystem.

ClawHub's open skill distribution model faces supply chain attack risks similar to those in software package ecosystems such as npm and PyPI. Existing records show that 335 malicious skills have appeared on ClawHub, often using professional documentation and harmless-looking names. A skill's \texttt{SKILL.md} file defines its purpose, trigger conditions, and available operations. When a skill is loaded, its definition is incorporated into the agent's context and influences the model's reasoning process. A malicious skill can embed adversarial instructions in \texttt{SKILL.md}, use execution privileges to access local files or network resources and transmit sensitive information externally, or interact with other legitimate skills to expand the scope of malicious behavior. The MCP protocol that supports OpenClaw's extensibility has itself become a security concern. Wang et al.~\cite{wang2025mcptox} reported that attacks against tool-using agents based on MCP achieved a success rate as high as 72.8\%.

The risk in the skill supply chain does not come only from external malicious actors. We observed that OpenClaw recently underwent a large architectural refactoring driven by AI-assisted programming, and that this refactoring caused a large number of existing skill interfaces to fail in batches. For users whose workflows depend on specific skills, a partially broken skill may silently return incomplete or incorrect results without reporting an error, and these results may then be used as the basis for later decisions in the agent's reasoning and action loop, thereby triggering the cascade effect discussed above. This event also exposed the absence of versioned API contracts in the OpenClaw skill ecosystem. More than 5,700 third-party skill developers built their skills against the current interfaces, but the platform did not provide guarantees of backward compatibility, advance notice mechanisms for interface changes, or transition periods for version migration. Development in the style of ``vibe coding'' often emphasizes fast iteration and functional delivery, while systematic evaluation of downstream compatibility impact may be weakened. For a platform with a large third-party skill ecosystem, changes to core interfaces are not merely technical decisions but also alter a trust commitment made to thousands of downstream developers and users. This event shows that for agent platforms that depend heavily on third-party skill ecosystems, platform stability is itself a security property.

\subsubsection{Multi-Session Contagion and Trust Boundaries}

OpenClaw's gateway layer manages multiple communication channels and sessions, and its persistent session mechanism allows information from earlier interactions to affect later behavior. The combination of these two architectural features means that security risk is no longer confined to a single session, but may spread across sessions and across users.

OpenClaw's persistent session mechanism maintains contextual continuity across interactions through session keys. If adversarial content from one interaction is persisted, for example if a malicious instruction inserted through indirect injection is saved as a user preference or a task record, it may continue to influence the agent's behavior in future interactions even after the original attack source no longer exists. Building on the attack scenario described in Section~3.2.1, if the state of a compromised session is persisted, adversarial influence may spread when the same user later initiates a follow-up interaction through another channel. If OpenClaw is then used to generate and send messages into channels involving other users, the contaminated output may create chained propagation across the platform.

Research confirms that such propagation is realistic. Gu et al.~\cite{gu2024agent} showed that in a system containing up to one million agents, injecting a single adversarial image into the memory of one agent can eventually infect all agents through normal pairwise interactions. Cohen et al.~\cite{cohen2024ai} demonstrated a self-replicating AI worm that spread across interconnected generative AI applications by poisoning RAG databases, with each compromised client able to infect 20 new clients within one to three days. Lupinacci et al.~\cite{lupinacci2025dark} found that across 18 state-of-the-art LLMs, 100\% could be compromised through inter-agent trust exploitation, executing malicious requests from peer agents that they would have rejected from human users. This reveals a key blind spot in current safety training, which is mainly designed for human-to-AI interactions rather than AI-to-AI interactions. Peign\'{e}-Lefebvre et al.~\cite{peignelefebvre2025multi} further identified a fundamental security-collaboration tradeoff: defenses intended to reduce adversarial propagation also reduce legitimate collaboration. This tradeoff is especially sharp in OpenClaw, because the platform's value is built on seamless collaboration across channels and sessions, while the information flow required for that collaboration is exactly the carrier of adversarial propagation.

\subsection{Defense Assessment}

The trajectory-based security audit of OpenClaw by Chen et al.~\cite{chen2026trajectory} provides integrated empirical support for the risks discussed above. The audit evaluated 34 representative test cases across six risk dimensions and found an overall pass rate of only 58.9\%. The 100\% pass rate in the hallucination and reliability dimension shows that the agent does not fabricate nonexistent tools or information. However, the 57\% pass rate for prompt injection and the 0\% pass rate for intent misunderstanding reveal a critical asymmetry: the system performs much worse at avoiding harmful actions than at avoiding fabricated facts. For pure text generation systems, hallucination is the main risk; for action-executing agents, incorrect actions and their irreversible consequences are the more serious threat.

The combined evidence from this section leads to the following assessment of the current defense posture.

First, the attack surface spans all architectural layers. From channel access at the gateway, to context construction in the runtime, to tool execution, to skill extension, and to persistent state management, every component introduces specific security risks that amplify one another.

Second, current defense mechanisms are seriously inadequate. The 17\% native defense rate reported by Ying et al.~\cite{ying2026uncovering}, the 58.9\% overall audit pass rate, and the systematic bypassing of 12 defense methods with over 90\% attack success rates~\cite{nasr2025attacker} together indicate that no single defense mechanism, whether training-based, detection-based, or prompt-engineering-based, is sufficient to address these threats.

Third, non-adversarial risks are equally serious. The platform refactoring event and the instruction amnesia incident show that the platform's own development practices and internal failures can lead to severe consequences even without a malicious attacker.

Fourth, architectural defenses appear to be the most promising direction. An emerging consensus in the research community~\cite{debenedetti2025camel,cheng2026openclaw} points toward combining privilege separation, sandboxed execution, and mandatory human-in-the-loop approval for irreversible actions. However, such strategies require deep architectural changes and continuous evaluation against adaptive attackers, and the security-collaboration tradeoff identified by Peign\'{e}-Lefebvre et al.~\cite{peignelefebvre2025multi} demands careful balance between protection and usability.

\section{OpenClaw vs. privacy}

% 4. OpenClaw vs. privacy

In this section, we examine the privacy implications of OpenClaw as a self-hosted, execution-capable AI agent. Unlike centralized chatbot services, OpenClaw is designed to operate under the deployer's control and may integrate persistent memory, local workspace files, external tools, messaging channels, and account credentials \cite{openclaw_trust}. This architectural model alters the locus and character of privacy risk. Rather than being governed primarily by the centralized data practices of a single provider, privacy in OpenClaw is shaped to a large extent by deployment choices, including system configuration, enabled tools and plugins, accessible data sources, and the presence or absence of isolation and auditing controls \cite{openclaw_gateway_security}. Consequently, although OpenClaw may alleviate certain privacy concerns associated with provider-side data governance in cloud-hosted AI services, it also introduces new privacy risks related to local privilege concentration, persistent storage, cross-context aggregation, and plugin supply chains \cite{openclaw_gateway_security,microsoft_openclaw_security_2026}. The following subsections therefore examine, in turn, OpenClaw's privacy boundaries and architecture, the principal privacy risks arising from this design, and the safeguards required for privacy-preserving deployment and use.

\subsection{Privacy boundaries and privacy architecture of OpenClaw}

Privacy boundaries refer to the practical limits on what data an AI system can access, retain, combine, and disclose. In OpenClaw, privacy boundaries are broader than in a conventional chatbot because the agent may draw not only on the current prompt, but also on resources beyond the immediate conversation. For instance, a single task may involve information stored in memory files, data from a linked email or calendar account, or actions performed through tools with access to local files. Accordingly, the relevant privacy question is not limited to what the user explicitly requests, but extends to what information the agent can in fact traverse across memory, credentials, tools, sessions, and channels during task execution. 

Consider a seemingly narrow request as figure~\ref{fig:openclaw_privacy_boundary} shown: a user asks OpenClaw only to reply to one email. On the surface, the task appears to require access only to that message. In practice, however, a broadly privileged deployment may also expose mailbox history, contacts, calendar entries, local files, stored preferences in memory, and context from other connected channels. The apparent scope of the task is therefore narrower than the system's actual scope of accessible data. Privacy should thus be understood first as a boundary problem: in OpenClaw, the relevant boundary is defined less by the user's immediate instruction than by the system's effective scope of accessible context.

\begin{center}
    \includegraphics[width=0.70\linewidth]{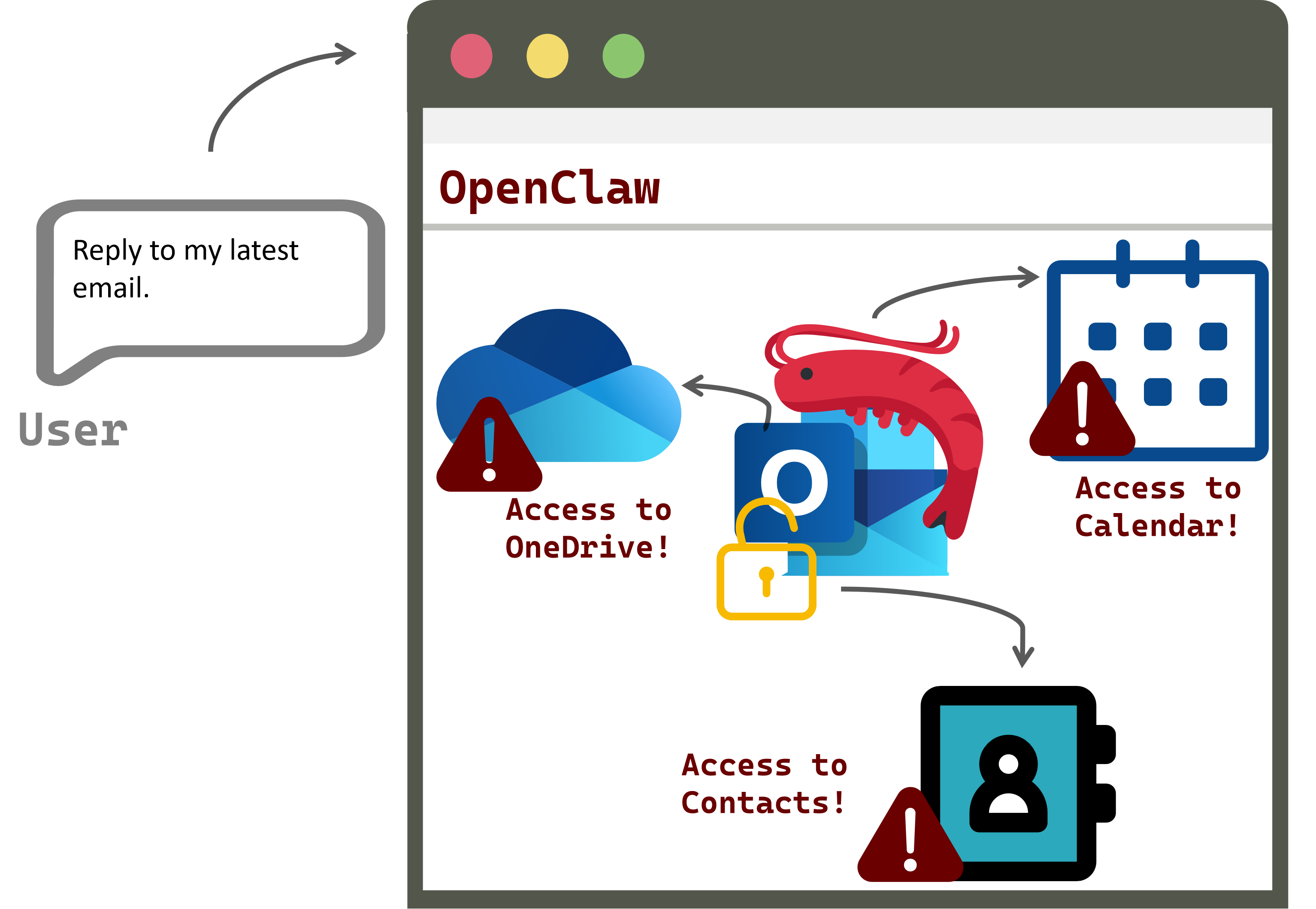}
    \captionof{figure}{Privacy risks in OpenClaw task execution: unintended expansion of data access across user-connected services.}
    \label{fig:openclaw_privacy_boundary}
\end{center}

If privacy boundaries describe the scope of possible exposure, privacy architecture describes the system design that produces and governs that scope. In this paper, privacy architecture refers to the structural arrangement through which OpenClaw accesses, stores, isolates, and propagates private data.According to OpenClaw's official documentation, this architecture is shaped by self-hosting, persistent memory, workspace-based storage, tool permissions, plugin mechanisms, session handling, and credential management \cite{openclaw_trust,openclaw_memory,openclaw_gateway_security,openclaw_plugin_architecture,openclaw_multi_agent}.

This architectural perspective is necessary because privacy in OpenClaw is not determined by hosting location alone. Official documentation notes, for example, that memory is stored as plain Markdown files in the workspace, that files on disk serve as the source of truth, and that agent directories, authentication profiles, and session isolation require careful separation and configuration. These are privacy-relevant architectural choices rather than incidental implementation details, since they determine where sensitive data reside, how long they persist, and how easily they may cross contexts. Therefore, privacy in OpenClaw should be understood as an architectural and governance issue rather than a simple consequence of local deployment, since its privacy posture depends on how storage, permissions, isolation, and persistence are organized in practice. This in turn gives rise to a distinct set of privacy risks.

\subsection{Privacy risks in OpenClaw}

OpenClaw introduces privacy risks because it can access, retain, combine, and reuse information beyond the immediate task context. The central privacy issue is not only whether the system can be attacked, but whether it can reach or preserve more personal data than users reasonably expect. In practice, these risks arise through persistent storage, over-broad data access, weak separation across users or contexts, plugin-mediated exposure, privacy disclosure triggered by untrusted input, and limited visibility into what data the agent has actually touched.

\subsubsection{Persistent memory and workspace storage.}
A major privacy risk in OpenClaw comes from the fact that information may remain on disk long after a single interaction has ended. Official documentation notes that memory can be written into workspace files such as daily notes and long-term memory files like \texttt{MEMORY.md}, which means that what the agent remembers is not only a temporary conversational context but a durable local record \cite{openclaw_memory,openclaw_gateway_security}.

The privacy problem here is accumulation. A user may separately ask OpenClaw to help schedule a doctor visit, remember food preferences for an event, and draft messages in a particular tone. None of these requests looks especially sensitive on its own. However, if the system stores all of them over time, it may gradually build a detailed record of the user's routines, health-related activities, social habits, and communication style. The resulting profile may be far more revealing than any single interaction. Even without an attacker, long-term retention increases the chance that personal information will later appear in the wrong context, be exposed through local compromise, or be reused in ways the user did not expect.

\subsubsection{Over-broad tool and account access.}
OpenClaw is designed to complete real tasks, which often requires access to email, calendars, files, browsing, messaging, and other connected services. The privacy concern here is not unsafe execution as such, but unnecessary data reach. A narrow request can open a much broader data path than the user realizes \cite{openclaw_trust,openclaw_gateway_security}.

A concrete example is a user asking OpenClaw to reply to one email. From the user's perspective, the task concerns a single message. In practice, a broadly privileged deployment may also expose the rest of the mailbox, contact records, past conversations with the same sender, calendar entries related to the topic, and even local files that help draft the reply. This matters even when the system does not explicitly quote private content back to the user. The surrounding metadata alone may reveal who the user communicates with, how often, at what times, and in what patterns. Over time, such access can support inference about work routines, personal relationships, or periods of absence. The privacy risk is therefore created by excessive data reach rather than by the visible wording of the task.

\subsubsection{Multi-user and cross-context interaction.}
OpenClaw also creates privacy risks when one agent instance is used across multiple people or contexts. Official documentation notes that direct messages may share the main session for continuity unless stronger isolation is configured, and also notes that session isolation does not by itself solve the broader host or operator trust problem \cite{openclaw_gateway_security}. This means that privacy boundaries in OpenClaw are often weaker than the boundaries users assume from the visible conversation alone.

A shared deployment makes this risk easy to see. Imagine a lab environment in which one student asks OpenClaw to remember that they will be away Friday afternoon for a medical appointment. Later, another student asks the same agent to summarize Friday availability or help schedule a meeting. If the deployment keeps context too broad, the first student's private information may shape the second interaction even though the two users did not intend to share context. The privacy harm here does not depend on an attacker. It arises from ordinary use when the separation between users, tasks, and contexts is weaker than the separation users reasonably expect.

\subsubsection{Plugins, skills, and supply-chain trust.}
Plugins and skills create a further privacy risk because they can enlarge what the system is able to see and process. Official plugin documentation states that native plugins run in the same process as the gateway and are not sandboxed \cite{openclaw_plugin_architecture}. In privacy terms, this means that installing a plugin is not just adding a new function. It may also expand the set of memory, session data, configuration files, and connected services that become reachable during execution.

A concrete example is a plugin advertised as helping with file organization or communication summaries. A user may install it because it appears useful and because the visible task seems routine. Once running, however, that plugin may be able to read stored memory, inspect session history, or access connected services while performing its helper function. Even if the plugin is not overtly malicious, an overly broad implementation may still collect or expose more data than the task requires. The privacy problem therefore lies not only in whether a plugin is intentionally malicious, but in whether users can realistically judge what data it will touch once it becomes part of the execution path. For this reason, plugin provenance, installation review, and restriction of extension paths are privacy controls as much as security controls \cite{openclaw_plugin_architecture,openclaw_gateway_security}.

\subsubsection{Untrusted input and privacy disclosure.}
OpenClaw may also create privacy harm when external content changes what information the agent chooses to retrieve, combine, or reveal. The security section already discusses indirect prompt injection as an attack mechanism. From a privacy perspective, the key issue is different. Even if the host remains intact, untrusted content may still redirect otherwise legitimate access toward unintended disclosure of personal data \cite{openclaw_gateway_security}.

Consider a user who asks OpenClaw to summarize recent communications about a project. During this process, the agent may read messages from Slack, email, attached files, and stored notes. If one retrieved source contains manipulative instructions, the privacy consequence may be that the agent includes unrelated personal details, pulls in information from another connected account, or sends a summary to a broader audience than the user expected. The task still looks ordinary, which makes the resulting privacy failure difficult for the user to notice or explain.

\subsubsection{Transparency and privacy auditing.}
A final privacy risk comes from limited visibility. OpenClaw gives operators more local control than a centralized service, but local control is not the same as practical transparency. Users and operators still need to know what data the agent accessed, which tools it invoked, what memory or files it updated, and whether any information left the local environment. If these steps are difficult to inspect, privacy exposure may remain hidden even when the system is self-hosted \cite{openclaw_context,openclaw_gateway_security}.

A simple example is a user asking OpenClaw to summarize recent communications. The user may think this requires only reading a few messages. In practice, the agent may read several channels, retrieve stored preferences, consult a connected account, and write new session state for later use. OpenClaw does provide controls such as session settings, tool restrictions, plugin loading rules, approvals, and \texttt{/context} summaries. However, these controls mainly help the operator configure the system. They do not always make it easy for an ordinary user to understand what was actually accessed, how long the data will remain available, or when it may later reappear in another task. Recent work on privacy auditing for agents similarly argues that effective oversight requires explicit records of sensitive data handling and policy compliance over time, rather than only static configuration \cite{zheng2025audagent,ukani2025browser_privacy}.

%4.3
\subsection{Privacy-preserving measures for OpenClaw}

The discussion above shows that privacy in OpenClaw depends less on local hosting itself than on how access, storage, extensions, and review are managed in practice. A privacy-preserving deployment should therefore restrict the agent's reachable context, separate trust domains, limit persistence of sensitive data, and make later inspection possible. In this sense, privacy protection is a deployment and architecture problem rather than a simple hosting choice.

\subsubsection{Minimize permissions and reachable context.}
A first principle is least privilege. OpenClaw's official security guidance recommends restricting inbound access through pairing and allowlists, and limiting high-risk tools such as \texttt{exec}, browser control, and network-facing tools to trusted agents only \cite{openclaw_gateway_security}. This is directly relevant to privacy because broad permissions enlarge the amount of personal data the agent can retrieve from connected accounts, files, and local applications. Zhou et al. similarly show that both users and models often disclose more context than is necessary, and argue for task-oriented data minimization \cite{zhou2025data_minimization}.

For OpenClaw, least privilege should be enforced at the task level. If a user asks the agent to reply to one email, the initial scope should cover only that message, the relevant thread, and the outbound reply action. It should not automatically include the full mailbox history, the contact list, the calendar, or unrelated local files. If broader access is later needed, the system should require explicit approval before expanding the task scope.

\subsubsection{Separate trust boundaries through dedicated runtimes and session isolation.}
OpenClaw documentation notes that per-user session isolation improves privacy but does not provide hostile-user isolation at the host or operator level \cite{openclaw_gateway_security}. For mixed-trust or multi-user settings, the safer pattern is to use separate gateways, or at minimum separate OS users, machines, or containers. Official guidance also recommends dedicated browser profiles and dedicated accounts for the runtime \cite{openclaw_gateway_security}. Karthikeyan et al. reach a similar conclusion in multi-agent settings and argue for structural separation of privacy boundaries \cite{karthikeyan2025agentcrypt}.

In practice, this means that users with different trust levels should not share the same workspace, browser profile, long-term memory files, or connected accounts. In a shared lab or household deployment, one common \texttt{MEMORY.md} file or one reused browser session can become a direct path for cross-user leakage even if chat sessions appear separate.

\subsubsection{Protect local memory, credentials, and retention paths.}
Because OpenClaw stores memory and other state on disk, privacy protection also requires strong storage discipline. Official documentation notes that memory is persisted in files such as \texttt{MEMORY.md} and daily notes, while security guidance warns that \texttt{\~/.openclaw/} may contain secrets, transcripts, credentials, and session data \cite{openclaw_memory,openclaw_gateway_security}. OpenClaw therefore recommends strict file permissions, dedicated state directories, and full-disk encryption on the gateway host \cite{openclaw_gateway_security}. Recent work further shows that agent memory can become a privacy boundary in its own right and may be abused through black-box extraction or unsafe retrieval \cite{wang2025memory_risks,sunil2026memory_poisoning}.

A privacy-conscious deployment should also distinguish between data types. Credentials and tokens should never be written into ordinary memory files. Task transcripts should be kept only as long as needed for debugging or review. Persistent memory should store only concise, approved summaries rather than raw sensitive details. For example, a writing preference may be retained, while a one-time code, temporary schedule, or private attachment should expire quickly or never enter long-term memory at all.

\subsubsection{Constrain plugins, external content, and code execution.}
Plugins, untrusted content, and tool execution should all be treated as privacy-relevant attack surfaces. OpenClaw's plugin architecture documentation states that native plugins run in-process with the Gateway and are not sandboxed, while the security documentation recommends explicit plugin allowlists, cautious installation practices, sandboxing for sensitive execution, and keeping unsafe external-content bypass flags disabled in production \cite{openclaw_plugin_architecture,openclaw_gateway_security}. Work such as \emph{MELON} and \emph{DRIFT} also supports a defense-in-depth approach to indirect prompt injection and unsafe tool use \cite{zhu2025melon,li2025drift}.

For OpenClaw, this means that plugins should be installed only from reviewed sources, privacy-sensitive deployments should avoid enabling native in-process plugins by default, and external webpages, pasted text, and attachments should be treated as untrusted input. A practical rule is that data retrieved from an untrusted source may be summarized for the user, but should not directly trigger file access, outbound messages, or browser actions without confirmation.

\subsubsection{Strengthen auditing, accountability, and privacy review.}
Finally, privacy protection requires governance as well as technical controls. OpenClaw provides policy surfaces such as configuration files, tool restrictions, channel controls, and context summaries, but these are useful only if they are actively reviewed \cite{openclaw_context,openclaw_gateway_security}. Zheng et al. propose \emph{AudAgent} for policy-based privacy auditing, while Ukani et al. show that browser agents expose a wide range of privacy weaknesses that require structured review \cite{zheng2025audagent,ukani2025browser_privacy}.

For OpenClaw, privacy auditing should record at least the triggering task, the channels and accounts touched, the tools invoked, the memory files read or written, and whether any information left the local environment. After completing a task, the system should make this access path reviewable by operators and, where appropriate, by users. In addition, deployments should support correction and deletion workflows so that retained memory can be inspected, revised, or removed when necessary.

Taken together, these measures suggest that privacy-preserving use of OpenClaw depends on disciplined deployment rather than on local hosting alone. A privacy-conscious configuration should combine task-bounded access, separated trust boundaries, protected local state, constrained extensions, and ongoing review of how personal data are stored, accessed, retained, and reused.

\section{OpenClaw vs. ethics}

In addition to security and privacy concerns, OpenClaw also raises important ethical issues. As a self-hosted, execution-capable AI agent, OpenClaw can access memory, local files, tools, plugins, and external services, and can therefore do more than merely generate text. In such systems, ethical concerns extend beyond output quality or model bias. More specifically, they include questions of how much decision-making is delegated to the agent, how user control is preserved, how responsibility is assigned when harm occurs, whether affected parties have meaningful awareness and consent, and how social harms may arise when agentic systems operate across shared contexts \cite{wef_agents_governance_2025,hahn2026ethical_agents}.

These concerns are especially important in OpenClaw because they arise from a concrete architecture rather than from abstract model capability alone. OpenClaw combines persistent memory, multi-channel communication, tool invocation, plugins, and long-running sessions in one operational system. As a result, the ethical stakes are shaped not only by what the model says, but also by what it can remember, retrieve, combine, and do on behalf of a user over time.

\subsection{Over-delegation, reduced user control, and cognitive debt}

A key ethical concern in agentic systems is that convenience can gradually turn into excessive delegation. OpenClaw allows users not only to ask for information, but also to hand over multi-step tasks such as organizing messages, prioritizing follow-ups, retrieving files, and triggering external actions. As a result, the user's role shifts from directly carrying out tasks to supervising the system. Over time, however, this supervision may weaken as users become accustomed to relying on the agent's summaries, recommendations, and actions. This concern is closely related to what Kosmyna et al.\ describe as \emph{cognitive debt}, namely the accumulation of delayed cognitive costs when mental effort is repeatedly offloaded to an AI assistant \cite{kosmyna2025cognitive_debt}.

In OpenClaw, this risk is reinforced by the combination of persistent memory, multi-channel message access, and action-capable tools. A user may begin by asking the agent only to summarize Slack updates, rank email follow-ups, or prepare routine replies. Yet the same system can retain earlier preferences, reuse prior context, and carry those judgments into later tasks. The ethical problem is therefore not only that the agent may make mistakes, but also that users may gradually lose sight of what data are being accessed, how decisions are being made, and when they should step in.

For example, a user may ask OpenClaw to ``summarize today's messages and decide what I should reply to first.'' Although the request appears simple, it transfers judgments about relevance, urgency, and social priority from the user to the system. In OpenClaw, this delegation is implemented through a concrete architectural pathway: the agent can aggregate messages from connected communication channels, combine them with preferences stored in persistent sessions, retrieve related project context through plugins or skills, and prepare replies through external communication tools. During repeated use, these interactions may also update the system's representation of the user's preferences, so that earlier choices continue to shape later priorities. As a result, OpenClaw is not merely producing a textual summary of an inbox. It can construct a prioritized representation of the user's social and professional obligations across multiple contexts. The agent may decide which messages are surfaced, which contacts are treated as urgent, which requests are delayed, and which replies are prepared for review or execution. If this becomes routine, users may stop checking intermediate steps and rely on the agent's priorities without sufficient review. In this way, delegation can quietly expand beyond what the user fully understands or intends, allowing OpenClaw to develop a quasi-independent decision profile that appears to act on the user's behalf while increasingly reflecting its own accumulated priorities, past inferences, and operational shortcuts. To avoid this, delegation should remain visible, reviewable, and easy to reverse, rather than gradually expanding through convenience and habit.

\subsection{Responsibility diffusion and accountability gaps}

OpenClaw also raises ethical problems of accountability because its actions emerge from a distributed chain of user instructions, model reasoning, tool execution, plugin behavior, and external services. In such systems, harms do not always arise from a single clearly attributable decision. Recent ethics scholarship on LLM-based systems describes this as a form of responsibility gap or ``many hands'' problem, in which moral and practical responsibility becomes fragmented across multiple actors and components \cite{constantinescu2025many_hands,lange2025accountability_agents}.

This issue is particularly visible in OpenClaw because a single outcome may depend on several architectural components at once. A wrong action may reflect how the runtime constructed context, what files or messages were retrieved, what a plugin returned, and what tools were permitted to execute. A harmful result can therefore emerge even when no single component appears solely responsible. The ethical problem is not merely that the system can fail, but that its mode of failure makes responsibility difficult to assign.

A practical example is a user asking OpenClaw to ``organize project materials and send an update.'' The agent may retrieve documents, generate a summary, choose recipients, and send the result through an external messaging or email tool. If confidential information is included by mistake, or if the message is sent to the wrong group, it may be unclear whether responsibility lies with the user's vague instruction, the agent's planning, the plugin used for retrieval, or the system configuration that permitted the action. The ethical issue here is not merely technical error, but the difficulty of identifying who should answer for the harm and what form of recourse is available to affected parties \cite{wef_agents_governance_2025,hahn2026ethical_agents}.

\subsection{Asymmetric consent and multi-user impact}

A further ethical concern arises when one person's authorization exposes other people's data, routines, or communications to agentic processing. In many real-world environments, information is shared by default: conversations involve multiple participants, documents are collaboratively edited, and messages circulate across organizational or social groups. Yet the decision to connect an agent to such environments is often made by a single deployer or account holder. This creates a form of asymmetric consent, in which one person enables the system while many others are affected by it \cite{oecd_agentic_landscape_2026}.

In OpenClaw, this concern is strengthened by cross-channel aggregation and persistence. One user may authorize the system to access a shared Slack workspace, email account, or team calendar, but the agent can then combine messages, schedules, and earlier context in ways that affect many others who did not directly consent to agentic processing. The ethical issue is therefore not only whether the initial access is technically authorized, but whether the later aggregation and reuse of that information are socially legitimate.

For example, an employee may connect OpenClaw to a shared team workspace and ask it to summarize discussions, extract action items, and draft follow-up messages. Even if the agent operates within that employee's formal permissions, other team members may be unaware that their comments, scheduling patterns, or working relationships are being aggregated and operationalized by an AI system. In OpenClaw, this concern is tied to the gap between configuration boundaries and consent boundaries: one user's connection of a channel or tool can make other people's shared communications reachable to agentic processing. In a similar way, a shared deployment in a lab or household may allow one user's preferences or reminders to become reachable in another person's later interaction if context boundaries are weak. This risk is particularly relevant to OpenClaw because its persistent sessions can preserve prior user-specific preferences, reminders, and interaction histories, which may later be reused outside the context in which they were originally provided. The ethical problem is therefore not only whether access is permitted, but whether all affected participants have meaningful awareness, voice, or recourse. In shared digital environments, responsible deployment requires that consent and governance reflect collective impact rather than unilateral enablement \cite{hahn2026ethical_agents,wef_agents_governance_2025}.

\subsection{Manipulation, social pressure, and relational harms}

Because OpenClaw is designed to act, not merely to respond, it is also vulnerable to ethically significant forms of manipulation. Recent reporting on research involving OpenClaw agents suggests that such systems can be induced through emotional pressure, guilt-tripping, or adversarial persuasion to reveal information, disable protections, or continue harmful behavior loops \cite{wired_openclaw_guilt_trip_2026}. This indicates that agentic systems may be socially manipulated in ways that are not captured by traditional concerns about false output alone.

For OpenClaw, the ethical significance of this problem is amplified by its ability to operate across communication channels while retaining contextual continuity. An attacker who appears in one channel as a colleague, teammate, or urgent contact may be able to influence the agent to retrieve information, draft a message, or continue an unsafe workflow in another connected context. In such a case, the manipulation does not remain confined to one exchange. It becomes part of a broader chain of delegated action.

The harm is therefore relational as well as technical. A user may trust OpenClaw to act in their interest, while another person manipulates that same system under the appearance of cooperation or urgency. For instance, an attacker might frame a request as helping a colleague, supporting a team goal, or resolving an emergency, thereby steering the agent toward actions that the user did not meaningfully authorize. In such cases, the harm is not only technical compromise, but also the exploitation of trust, the distortion of delegated agency, and the erosion of confidence in human--AI interaction \cite{toward_safe_responsible_agents_2026,lange2025accountability_agents}.

Overall, the ethical challenges of OpenClaw arise not simply from what the system can say, but from what it can remember, infer, and do across contexts. They concern how agency is redistributed, how responsibility is assigned, whose consent is counted, and how harms emerge when autonomous or semi-autonomous systems act within shared social environments. Ethical evaluation of OpenClaw should therefore focus not only on model behavior, but also on the human, organizational, and relational consequences of delegated AI action.
% 1. AI 代理伦理：OpenClaw 的伦理问题不再只在于生成内容本身，而在于系统已经开始代替人执行真实操作。

% 2. 人类能动性与控制边界：当 agent 持续记忆、推断并调用工具时，帮助用户与替代用户之间的界线会变得模糊，人的知情决策和自主控制也可能被削弱。

% 3. 问责与透明性：一旦 OpenClaw 做出不当行动，责任可能分散在用户、系统、插件和外部输入之间，同时普通用户也很难真正看清系统访问了什么、保存了什么、为何这样行动。

% 4. 授权、同意与多用户影响：当系统访问文件、消息、凭据或共享空间时，受影响的人未必都真正同意，多用户和共享环境会让这个问题更复杂。

% 5. 法律与治理挑战：OpenClaw 的本地执行、外部集成和半自主行为，使现有法律、责任划分和治理框架未必足以应对实际后果。

% 6. 更广泛的社会影响：随着这类 agent 进入工作和日常数字环境，依赖加深、监督减弱、能力退化以及控制关系变化都可能成为新的伦理问题。

\section{Reliability and Traceability in OpenClaw}

The preceding discussion has shown that OpenClaw raises substantial challenges in security, privacy, and ethics. For an execution-capable agent, however, it is also necessary to ask two further questions: whether the system behaves dependably in practice, and whether its behaviour can be reconstructed when failures occur. These questions correspond to reliability and traceability. Reliability concerns whether the agent can perform tasks consistently and safely across repeated or changing conditions, while traceability concerns whether its decisions and actions can be reconstructed after the fact. Together, they provide a basis for evaluating not only what risks OpenClaw may create, but also whether those risks can be detected, analysed, and reduced. As shown in Figure~\ref{fig:openclaw_reliability_traceability}, these two properties apply to the same execution chain from user request to final action: reliability concerns whether the chain performs dependably, whereas traceability concerns whether the chain can be inspected and reconstructed afterwards.

\begin{center}
    \includegraphics[width=0.70\linewidth]{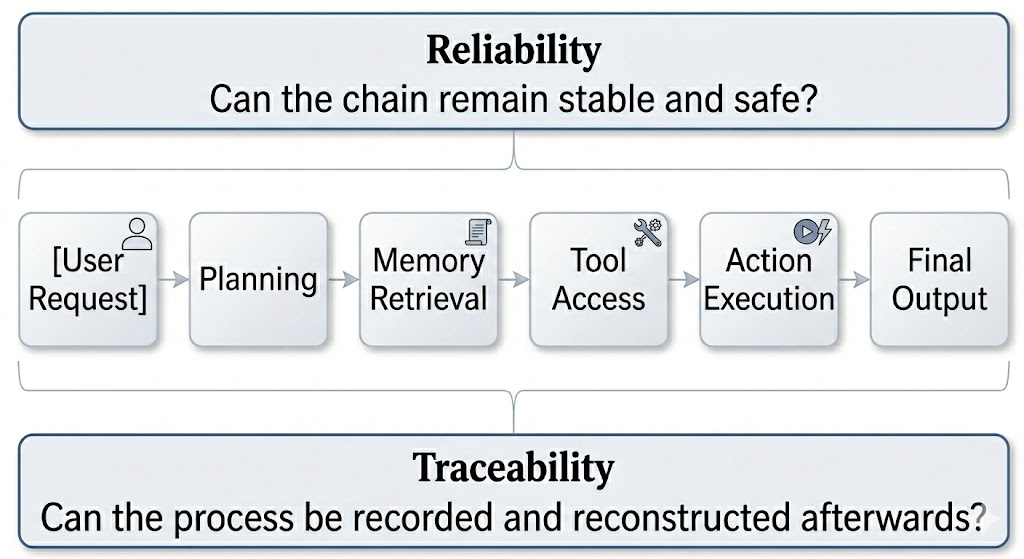}
    \captionof{figure}{Reliability and Traceability of the OpenClaw Execution Chain.}
    \label{fig:openclaw_reliability_traceability}
\end{center}

\subsection{Reliability in OpenClaw}

Reliability refers to the extent to which an agent can complete tasks consistently, robustly, and safely across repeated runs and changing execution conditions. For LLM-based agents, reliability cannot be captured well by a single successful run, because behaviour may vary across repeated executions, semantically equivalent task variants, and infrastructure disturbances such as timeouts or rate limits. Recent work has therefore argued that agent reliability should be evaluated in terms of repeated-execution consistency, robustness under perturbation, and tolerance to execution faults rather than only one-shot task success \cite{gupta2026reliabilitybench,science_agent_reliability_2026}.

In OpenClaw, reliability is especially important because errors can propagate across multiple steps and may lead to irreversible actions. A misunderstanding in an early step may affect later file access, tool invocation, or message sending, so the issue is not merely whether one answer is correct, but whether the overall execution remains dependable. For example, a user may ask OpenClaw to retrieve project files, summarize them, and send an update message. Even if the task appears routine, small variations in context, tool availability, or early-step interpretation may lead to different execution paths and different outcomes across runs. In this sense, reliability in OpenClaw means whether the agent can be trusted to perform a task in a stable, repeatable, and safe manner when operating over memory, tools, and persistent context.

\subsection{Traceability in OpenClaw}

Traceability is the ability to reconstruct how an agent's output or action was produced, including the relevant inputs, intermediate decisions, memory use, and tool calls. In agent systems, this usually requires structured records of execution rather than only final outputs. Recent work on provenance and auditability argues that agentic systems need fine-grained execution traces that link prompts, decisions, actions, retrieved evidence, and downstream effects, so that failures can be examined and claims can be audited after the fact \cite{souza2025provagent,rasheed2026claim_auditability}. Related work on observability standards for agentic systems similarly emphasizes the need to trace tasks, actions, artifacts, memory, and agent interactions in a structured way \cite{otel_agentic_semconv_2025}.

For OpenClaw, traceability therefore means more than ordinary logging. It means that when the agent reads memory, retrieves files, invokes tools, or produces an action, the execution path should remain inspectable afterwards. If OpenClaw sends the wrong message, modifies the wrong file, or reaches an unsafe conclusion, traceability should make it possible to determine what context was used, which step introduced the problem, and how the final outcome was produced. In this sense, traceability concerns whether the agent's behaviour remains inspectable and reconstructable after execution.

Taken together, reliability and traceability are not secondary implementation details, but core properties of execution-capable agents. Reliability determines whether OpenClaw can be trusted to perform tasks in a stable and safe manner, while traceability determines whether failures can be meaningfully analysed and corrected. These properties therefore provide an important foundation for future research on robust evaluation, failure diagnosis, and auditable agent design.

\section{Future Works}

The analysis of OpenClaw suggests that future research on agent-based systems should move beyond isolated safeguards and toward system-level frameworks that jointly address execution, data reachability, responsibility, and verifiability. In particular, future work may be organized around four complementary directions: security, privacy, ethics, and traceability.

\subsection{Security}

\begin{itemize}
\item \textbf{Memory poisoning and persistent-state protection.}  
A major research direction is how to prevent malicious or misleading content from becoming part of OpenClaw's persistent state. Future work should focus on concrete memory objects such as session histories, daily notes, retrieved memory entries, and files like \texttt{MEMORY.md}. One practical direction is to label, quarantine, or expire newly written memory before it is reused in later planning or action. This would reduce the chance that a poisoned summary, reminder, or preference record silently becomes part of the agent's long-term context.

\item \textbf{Reliable sandboxing for execution-capable agents.}  
Another important direction is the design of sandboxing for the specific execution paths that make OpenClaw high risk in practice. Future work should separate file access, browser sessions, shell execution, plugin code, and outbound communication into different control domains. A practical direction is to ensure that high-risk tool combinations are blocked by default and that irreversible actions require stronger approval before execution.

    \item \textbf{Skill and tool auditing.}  
    A further research need is the systematic auditing of skills and tools. In agent platforms, risks do not arise only from the core model, but also from the permissions, behaviours, and interactions of reusable tools, plugins, and skills. Future work should develop auditing frameworks that evaluate what each skill or tool is allowed to access, how it behaves in realistic execution settings, and how combinations of components may create new attack surfaces or misuse opportunities.
\end{itemize}

\subsection{Privacy}

\begin{itemize}
    \item \textbf{Task-bounded data access.}  
    An important research direction is how to define and enforce task-bounded data access for agent systems. In multi-step execution, privacy risk depends not only on what permissions an agent has in principle, but also on what information becomes reachable in practice during planning, retrieval, and tool use. Future work should therefore study how to restrict agents to the minimum data required for a given task, while preventing unnecessary access to unrelated memory, files, accounts, or communication channels. For OpenClaw, this could be operationalized as a per-task reachability manifest that records which memories, files, channels, accounts, and tools are available for a specific request, and blocks components outside that declared scope unless the user explicitly expands it.

    \item \textbf{Privacy-aware memory design.}  
    Another key direction is the design of privacy-aware memory mechanisms for persistent agents. Since long-term memory can improve continuity and task performance, future systems need ways to balance utility against privacy risk. In OpenClaw, this suggests separating persistent session memory into typed stores, such as user preferences, task history, credentials, reminders, and shared-context records, with different retention periods, retrieval rules, and user-facing deletion controls for each type.

    \item \textbf{Control of privacy inference from cross-source aggregation.}  
    A further challenge is how to detect and limit privacy-sensitive inference created by cross-source aggregation. Agent systems can combine messages, documents, calendars, memory, and external tool outputs to infer information that was never explicitly disclosed in any single source. Future research should therefore explore methods to measure, surface, and constrain such inference, so that useful cross-source reasoning does not silently become a source of privacy leakage. For OpenClaw, one concrete direction is to flag high-risk aggregation paths, such as combining messages, calendars, files, and persistent memory to infer a person's availability, relationship, work pattern, or priority status, even when none of these attributes is explicitly stored in a single source.

\end{itemize}

\subsection{Ethics}

\begin{itemize}
\item \textbf{Designs that prevent convenience from becoming over-reliance.}  
A key ethical direction is to design OpenClaw so that convenience does not quietly become over-reliance. Future work should distinguish more clearly between summarizing, prioritizing, and acting, and require different levels of user involvement at each stage. Another practical direction is to show which memories, preferences, or past interactions shaped a recommendation, so that users can see when the agent is beginning to make decisions on their behalf.

    \item \textbf{Responsibility allocation and recourse mechanisms.}  
    Another important direction is the development of practical mechanisms for assigning responsibility and providing recourse when harms arise from agent-mediated actions. This includes methods for clarifying who authorized a task, how the agent reached a decision, which components were involved in execution, and how affected users can request correction, reversal, or review when something goes wrong.
\end{itemize}

\subsection{Reliability and Traceability}

\begin{itemize}
    \item \textbf{Repeated-run reliability evaluation.}  
    A practical research direction is to evaluate agent reliability beyond one-shot task success. Recent work such as \emph{ReliabilityBench} argues that agent reliability should be measured in terms of consistency across repeated runs, robustness under semantically equivalent perturbations, and tolerance to tool or API failures, rather than only single-run accuracy \cite{gupta2026reliabilitybench}. Future work should therefore develop OpenClaw-specific benchmarks that repeatedly run the same task under semantically equivalent prompts, changed tool responses, permission changes, or partially unavailable plugins, and measure whether the selected tools, accessed data sources, final actions, and safety outcomes remain stable.

    \item \textbf{Fine-grained provenance for multi-step execution.}  
    Another practical direction is to design structured provenance for agent workflows. Recent work such as \emph{PROV-AGENT} and \emph{Claim-Level Auditability for Deep Research Agents} shows that agent systems need fine-grained traces linking prompts, intermediate decisions, retrieved evidence, tool calls, and downstream effects, so that failures can be reconstructed and audited after the fact \cite{souza2025provagent,rasheed2026claim_auditability}. Future work should therefore explore how OpenClaw can record execution provenance in a form that is both machine-processable and useful for debugging and post-hoc analysis. Such provenance could be implemented as a structured execution trace that links each user instruction to the retrieved context, memory entries, selected skills, tool calls, permission checks, generated outputs, and external side effects.
\end{itemize}

\section{Conclusion}
This paper systematically analyzes OpenClaw as an execution-capable autonomous AI agent platform, examining its system architecture and identifying the security, privacy, and ethical risks arising from its integration of planning, memory, and tool use. It highlights how factors such as prompt injection, over-privileged access, persistent state, and plugin ecosystems interact and amplify risks in real multi-step execution, and shows that local deployment does not inherently guarantee privacy due to expanded data access boundaries. The paper further discusses shifts in user control, accountability, and consent under delegation-based interaction. Overall, it argues that the key challenges of agent systems stem from the coupling between architecture, deployment, and usage, and outlines directions for improving permission control, isolation, traceability, and governance. More broadly, we call for the development of agentic AI systems that are not only capable and efficient, but also secure by design, privacy-preserving in operation, and ethically aligned with human interests and social expectations.

\paragraph{Note on references.}
Because research on agentic AI security remains a rapidly developing area, some recent works cited in this paper are currently available as arXiv preprints and may not yet have undergone peer review. Where published versions were available, we cited the latest formal publication.

\bibliographystyle{elsarticle-num}
%\bibliographystyle{cas-model2-names}

\bibliography{cas-refs}



\end{document}